\documentclass[letterpaper,aps,prd,onecolumn,groupedaddress,preprintnumbers,superscriptaddress,showpacs,amsmath,amssymb,nofootinbib]{revtex4}

\usepackage{amsmath,amssymb,latexsym,graphicx}
\usepackage{dcolumn}    
\usepackage{color}

\definecolor{red}{rgb}{1,0,0}
\definecolor{blue}{rgb}{0,0,1}
\definecolor{green}{rgb}{0,1,0}

\begin{document}

\title{ Origin of black string instability }

\author{  Hideaki Kudoh  }
\email{kudoh_atmark_utap.phys.s.u-tokyo.ac.jp}
\affiliation{
  Department of Physics, The University of Tokyo, Tokyo 113-0033, Japan   
}

\begin{abstract}
It is argued that many nonextremal black branes exhibit a classical Gregory-Laflamme (GL) instability. Why does the universal instability exist? 
To find an answer to this question and explore other possible instabilities, we study stability of black strings for all possible types of gravitational perturbation. 
The perturbations are classified into tensor-, vector-, and scalar-types, according to their behavior on the spherical section of the background metric. The vector and scalar perturbations have exceptional multipole moments, and we have paid particular attention to them. 
It is shown that for each type of perturbations there is no normalizable negative (unstable) modes, apart from the exceptional mode known as $s$-wave perturbation which is exactly the GL mode.
We discuss the origin of instability and comment on the implication for the correlated-stability conjecture.
\end{abstract}

\preprint{UTAP-550}
\preprint{NSF-KITP-06-05}
\preprint{hep-th/0602001}


\maketitle
 
    \section{Introduction}

Stability of a given spacetime is a crucial issue from many standpoints. 
In general relativity, a stable spacetime will be realized by a dynamical evolution starting from a generic set of initial data on a Cauchy surface. 
However stability in general relativity is frequently subtle issue, and because of that it becomes important and interesting in its own right. 
From a string theory perspective, it is interesting to know what spacetimes are appropriate backgrounds for studying string propagation and its dynamics. 
Besides, information of gravitational dynamics and properties are useful to understand Yang-Mills theory by means of gauge/gravity dualities, and vice versa~\cite{Hubeny:2002xn,Maeda:2005cr,Aharony:2004ig,Maldacena:2001kr,Hawking:2005kf,Witten:1998zw,Harmark:2004ws,Harmark:2005dt}. 
In this respect,
instability on the gravitational side is an indicator of interesting
gauge theory dynamics, such as phase transition and
so on~\cite{Horowitz:2005vp}.

The fundamental generic instability is Gregory-Laflamme (GL) instability~\cite{Gregory:1993vy}, which is accompanied by a uniformly smeared horizon. 
The fundamental phenomena is however one of long-standing puzzles in gravity.
For example, 
(i) {\it{what is the necessary and sufficient condition for the onset of a dynamical instability of a horizon?}}
(ii) {\it{Why is a uniform horizon unstable?}} 
The first question was addressed by a so-called correlated-stability
conjecture (CSC)~\cite{Gubser:2000ec,Gubser:2000mm}. 
Namely, the onset of the dynamical instability of black brane will be the same as the onset of (local) thermodynamic instability. 
The second question is more fundamental and naive. The origin of the instability might have deep connection with quantum aspect of gravity, since the onset of instability is predictable by black hole thermodynamics due to CSC. 
Here we would like to pursue the question from classical aspect of gravity. 
(See \cite{Cardoso:2006ks} for fluid analogy of GL phenomena.)

First of all, we do not know full dynamics of unstable
black objects in higher dimensions \cite{Choptuik:2003qd}. 
In particular, as far as the present author knows, a complete analysis of (in)stability has not been carried out.   
(A numerical investigation for the 5-dimensional black string in the braneworld model with AdS bulk was performed in \cite{Seahra:2004fg}.)
In fact, even for the higher dimensional Schwarzschild black holes (BHs), its dynamical stability was established in recent years by Kodama and Ishibashi (KI)~\cite{Kodama:2003jz,Ishibashi:2003ap,Kodama:2003kk}. 
The instability found by Gregory and Laflamme is the $s$-wave
mode, and the perturbation is ``minimum'' deformation of horizon. For perturbations with higher multipole moments, similar instability might persist. An interesting point is that existence of instability implies existence of a critical static
mode and the mode could be continued to a state with nonperturbatively deformed horizon~\cite{Gubser:2001ac,Wiseman:2002zc,Kudoh:2004hs,Kudoh:2003ki,Sorkin:2003ka}, 
so that any extra instability implies extra static sequence of solutions.
Besides, they will have physical meaning in Euclidean space~\cite{Gross:1982cv,Allen:1984bp,Prestidge:1999uq,Sarbach:2004rm}. 
\footnote{
The stability argument in Ref. \cite{Gross:1982cv} is sometimes applied to black strings, using Wick rotation. 
The proof of stability for higher multipole moments assumes that all eigenvalues are real under the periodic Euclidean time. 
In general, this assumption for eigenvalues and boundary conditions is crucial for stability argument, and we should not naively apply the argument to discuss the stability of black strings. 
}
In addition to the stability issue of uncharged black branes, complete stability of BPS state with respect to {\it{all}} possible types of perturbations, which should include breaking of supersymmetry,  remains an open question, although there are several evidence for it~\cite{Gregory:1995tw,Hirayama:2002hn,Kang:2004hm}.

In order to promote greater understanding of the nature of black string/brane, it is inevitable to investigate the stability with respect to all the types of perturbations. Following to the general gauge-invariant formalism for higher dimensional maximally symmetric BHs by KI, we develop a general perturbation theory of black string and tackle the stability problem.
(See also \cite{Kodama:2000fa,Mukohyama:2000ui} for the basic work related to the gauge-invariant formalism of maximally symmetric spacetimes.)
In this approach, the perturbation variables are classified into three types, those of tensor, vector and scalar modes, according to the type of harmonic tensor used to expand the perturbation variables. 
Contrary to the perturbations for the maximally symmetric BHs, vector and scalar type perturbations will not have simple master variables due to extra physical degrees of freedom. 
We study stability of these perturbation variables.

The paper is organized as follows. In the next section we first classify perturbations into tensor, vector, and scalar-types with respect to the maximally symmetric $n$-dimensional spacetimes. Then for each type of perturbations, we express the Einstein equations in terms of them.
In Sec.~\ref{Stability analysis: Tensor and Vector}, we study stability of tensor and vector perturbations and found that there is no instability in these perturbations. 
In Sec.~\ref{sec:Scalar perturbations}, the stability analysis of scalar perturbation will be carried out. 
The unstable GL mode is an exceptional mode in the present perturbation scheme and we discuss that there is no other unstable mode in the black string perturbations. 
The origin of such an exceptional mode will be clarified in comparison with the perturbations for the maximally symmetric BHs.
Section~\ref{sec:Summary} is devoted to summary and discussion. 
Throughout this paper we follow the notation in Refs.~\cite{Kodama:2003jz,Ishibashi:2003ap,Kodama:2003kk}.

\section{General Perturbation Theory}

As our background spacetimes, we consider the $D=(n+3)$-dimensional metric of the form
\begin{eqnarray}
    d\bar{s}^2_D = \bar{g}_{AB} dx^A dx^B = g_{ab} dy^a dy^b+r^2d\sigma_n^2 + dz^2, 
\label{eq:ds^2}
\end{eqnarray}
where $g_{ab}$ is the Lorentzian metric of the two-dimensional orbit spacetime, and $d\sigma_n^2 = \gamma_{ij}(y) dy^i dy^j$ is the metric of the $n$-dimensional maximally symmetric space ${\cal K}^n$ with sectional curvature $K=0,\pm 1$.
Throughout this paper, we use the notation $a,b=0,1$,~~$i,j=2, \cdots, n+1$ and 
$\alpha,\beta\cdots=0, \cdots,  n+1$. The covariant derivative with respect to the metric $g_{ab}$ and $\gamma_{ij}$ are defined as $D_a$ and $\widehat{D}_i$, respectively.
In the followings, we develop a general perturbation scheme for black objects with co-dimension one. The perturbation will be specialized to the black string perturbations in the next section (See footnote \ref{footnote2} for more general perturbations.)

Most general metric perturbations $\delta g_{AB}$ for this background spacetimes are 
\begin{eqnarray}
    ds^2 = ( g_{AB}+ \delta g_{AB}  ) dx^A dx^B. 
\end{eqnarray}
Utilizing gauge degrees of freedom, $x_{A} \to x'_{A} = x_{A} + \xi_{A} (x_{\beta})$, we can eliminate perturbations in $z$ direction at any times, taking a Gaussian normal coordinates: 
\begin{eqnarray}
ds^2 = (g_{\alpha\beta} + \delta g_{\alpha\beta} ) dx^\alpha dx^\beta + dz^2 . 
\label{eq:dg_ab}
\end{eqnarray}
This gauge fixing is however not complete. 
There are two types of residual gauge degrees of freedom.
The corresponding infinitesimal coordinate transformations are
\begin{eqnarray}
    \xi_z = P(x_{\alpha}),\quad
    \xi_\alpha = - z \partial_\alpha P(x_{\beta})  , 
\label{eq:residual gauge PQ1}
\end{eqnarray}
and
\begin{eqnarray}
    \xi_z = 0,
\quad
    \xi_\alpha =  Q_{\alpha} (x_{\beta}).
\label{eq:residual gauge PQ2}
\end{eqnarray}
The first one corresponds to shifting $z=\mathrm{const.}$ surface. 
The second is the gauge transformation transverse to a $z=\mathrm{const.}$ surface, and hereafter we call this ``transverse" gauge degrees of freedom.

Because the background spacetimes are translationally invariant along $z$ direction, we can take arbitrary hypersurface of 
\begin{eqnarray}
z=\mathrm{const.} = z_c 
\end{eqnarray}
to study the perturbations without loss of generality.   
This approach is a sort of an effective theory approach.  
In this approach, the residual gauge $P$ is fixed once we take a $z=\mathrm{const.}$ surface, on which we will study perturbations.

At this point, if we consider only homogeneous perturbations along the $z$ direction, the general perturbations (\ref{eq:dg_ab}) 
is the same as the gravitational perturbations of maximally symmetric black holes in higher dimensions studied by Kodama and Ishibashi.
Their perturbation theory is most generic and based on gauge-invariant scheme, yielding master variables for each type of perturbations. 
Following to their perturbation theory, below, we develop
{\it{transversely}} gauge invariant perturbation theory, so that perturbation variables independent of the residual gauge (\ref{eq:residual gauge PQ2}) are introduced. 
A point is that the general perturbation provides transparent perturbation scheme, which can be directly compared with the perturbations of maximally symmetric black holes.
\footnote{
\label{footnote2}
We have decomposed the metric into $2\times n  \times 1$ space with employing the gauge fixing. 
More general formulation will be possible by decomposing the metric into $m \times n$ space~\cite{Kodama:2000fa}, where 
$m$ is $m \ge 3$ depending on the translationally invariant spatial dimensions of black brane.  
By employing such decomposition, we can use many covariant formulas for the higher dimensional maximally symmetric BHs in \cite{Kodama:2003jz} without significant changes, although such fully gauge invariant equations give more messy equations of motion for each variables.
In this picture, it is easy to count a number of physical degrees of freedom for each type of perturbation. 
The physical degrees of freedom for tensor and vector $F^a$ (see Eq.(\ref{eq: def vector F_a})), are $1\times [\mathbb{T}_{ij} ] $ and $(m-1) \times [\mathbb{V}_{i} ] $, respectively, taking into account the number of constraint equations for vector perturbation. 
Here $[\mathbb{T}_{ij} ] =(n +1)(n-2)/2$ and $[\mathbb{V}_{i} ] =(n-1)$ are the number of degrees of freedom for the respective harmonics. 
The physical degrees of freedom for scalar perturbation,  $F$ and $F_a^b$, are $(m^2 + m +2)/2 - (m+1) = m(m-1)/2$, subtracting the number of constraint equations for scalar perturbation.  
The total gravitational degrees of freedom are $(n+m)(m+n -3)/2$.
}

\subsection{Tensor-type perturbations}

We begin by considering tensor perturbations, which are given by
\begin{eqnarray}
  \delta g_{ab} = 0,
~~\delta g_{ai} = 0,
~~\delta g_{ij} = 2r^2 H_T \mathbb{T}_{ij},
~~\delta g_{\alpha z} = 0,
\label{eq:def tensor part}
\end{eqnarray}
where $H_T$ is a function of $\{ t,r,z \}$, and the harmonics tensors $\mathbb{T}_{ij}$ are defined as solutions to the eigenvalue problem on the $n$-sphere; 
\begin{eqnarray}
    (\widehat{\Delta}_n + k_T^2) \mathbb{T}_{ij} = 0, 
\quad
    {\mathbb{T}^i}_i = 0,
\quad
    \widehat{D}_j {\mathbb{T}^j}_i = 0. 
\label{eq:def tensor harmonics}
\end{eqnarray}
Here, $\widehat{D}_j$ is the covariant derivative with respect to the metric $\gamma_{ij}$ and  $\widehat{\Delta}_n \equiv \gamma^{ij} \widehat{D}_i \widehat{D}_j$. 
In these equations we have omitted the indexes labeling the harmonics and the summation over them. 
For $K=1$, the positive eigenvalue $k_T^2$ for a discrete set,
$ k_T^2 = l(l+n-1)-2,~~l=1,2,\cdots. $

The tensor perturbations are essentially transversely gauge invariant.
Following ~\cite{Kodama:2000fa,Kodama:2003jz,Ishibashi:2003ap,Kodama:2003kk}, we introduce a new  variable  $\Phi = r^{n/2} H_T$.  
The master equation follows from the vacuum Einstein equations (\ref{eq:delta R_ij}), 
\begin{eqnarray}
     \left\{  \square  - \frac{1}{r^2}
     \left[
          k_T^2 + 2n K
        + \frac{n-4}{2} ~ r \square r
        + \frac{n^2-10n +8}{4} (Dr)^2
     \right] \right\}\Phi + \Phi_{,zz}  = 0,
\label{eq:EOM tensor}
\end{eqnarray}
where $\square = D^a D_a$ denotes D' Alembertian operator in the two-dimensional orbit space. 
We remind that there is no tensor-type harmonics on a 2-sphere, so that the tensor perturbations only exist for $n \ge 3$.

\subsection{Vector-type perturbations}
Perturbations of the vector type can be expanded in terms of vector-type harmonic tensors $\mathbb{V}_i$ satisfying
\begin{eqnarray}
(\widehat{\Delta}_n + k_V^2 ) \mathbb{V}_i =0, 
\quad
\widehat{D}_j  \mathbb{V}^j = 0. 
\end{eqnarray}
As in the case of tensor-type harmonics, the eigenvalues $k_V^2$ are positive definite.
For $ K = 1 $ the eigenvalues form a discrete spectrum given by 
\begin{eqnarray}
    k_V^2 = l(l+n-1) - 1,~~l=1,2,\cdots. \quad \bigl( K=1 \bigr)
\end{eqnarray}  
In terms of vector harmonics, metric perturbations are expanded as  
\begin{eqnarray}
    \delta g_{ab} = 0,
    ~~\delta g_{ai} = rf_a {\mathbb{V}}_i,
    ~~\delta g_{ij} = 2r^2H_T {\mathbb{V}}_{ij},
    ~~\delta g_{\alpha z} = 0, 
\end{eqnarray}
where ${\mathbb{V}}_{ij}$ and ${\mathbb{V}}_{j}$ satisfy
\begin{eqnarray}
    {\mathbb{V}}_{ij} = -\frac{1}{2k_V}(\widehat{D}_i \mathbb{V}_j + \widehat{D}_j \mathbb{V}_i ),
\quad
    {{\mathbb{V}}^i}_i = 0, 
\quad 
    \widehat{D}_j {{\mathbb{V}}^j}_i = \frac{k_V^2-(n-1)K}{2k_V} {\mathbb{V}}_i. 
\end{eqnarray}
Note that ${\mathbb{V}}_{ij}$ also satisfy $ [{\widehat{\Delta}} + k_V^2 -(n+1)K] \mathbb{V}_{ij}=0$. 
The special mode $k_V^2 = (n-1) K$ is known as the exceptional mode for the vector perturbations, since ${\mathbb{V}}_{ij}$ vanishes for this mode.

For $k_V^2 \neq (n-1) K >0 $, transversely gauge invariant quantity is 
\begin{eqnarray}
    F_a (t,r,z)= f_a + r D_a \left(\frac{H_T}{k_V}\right). 
\label{eq: def vector F_a}
\end{eqnarray}
The vacuum Einstein equations, $\delta R_{ij}=0$ and $\delta R_{ai}=0$, reduces to
\begin{eqnarray}
&& D_a( r^{n-1} F^a ) 
  = - r^n  \frac{\partial^2  }{\partial z^2} \left(  \frac{ H_T}{k_V} \right), 
\label{eq:eq for vector F_a 1}
\\
&& 
D_a \left(  r^{n+1} 
      F^{(1)} \right)
 - m_V \, r^{n-1} \epsilon_{ab} F^b 
  =  - r^{n+1} \epsilon_{ab} \,  f^b _{~,zz}  ~~ . 
\label{eq:eq for vector F_a 2}
\end{eqnarray}
where $m_V \equiv k_V^2 - (n-1) K$, and we have introduced 
\begin{eqnarray}
    F^{(1)}  
   =   r~ \epsilon^{ab}  D_a \left( \frac{F_b}{r} \right). 
\label{eq:def F^(1)}
\end{eqnarray}
The Einstein equation $\delta R_{iz}=0$ gives a non-vanishing equation, but it is not an independent equation. 
Combining these two equation, we obtain an equation for $F_a$, 
\begin{eqnarray}
&& \epsilon^{ad}  D_d 
        \left[
            r^{n+2}  D_b \left(\frac{  \epsilon^{bc} F_c}{r}\right) 
        \right]
 + r^{n+2} D^{a} \left[ \frac{1}{r^n}D^c(r^{n-1}F_c) \right]
 - m_V \, r^{n-1}  F^a 
  =  -  r^{n+1}  F^a_{~,zz} . 
\label{eq:eq for vector F_a 3}
\end{eqnarray}
Therefore our stability problem is reduced to solve the equation of motion (EOM) for the vector $F_a$. 
The vector (\ref{eq: def vector F_a}) has been constructed to be invariant under the gauge transformation which is independent of $z$. 
Thus any solutions of the evolution equation (\ref{eq:eq for vector F_a 3}) have physical meaning.

Here we note that for the zero mode $\partial_z \partial_z  F_a=0$ the divergenceless condition (\ref{eq:eq for vector F_a 1}) holds for the vector $F_a$. 
From this condition, a master variable can be introduced, and the second equation (\ref{eq:eq for vector F_a 2}) with employing the master variable reduces the Regge-Wheeler equation for $n=2$, $K=1$. 
By contrast with the zero mode, the KK modes have one extra physical degree of freedom. 
The two physical modes are governed by Eq. (\ref{eq:eq for vector F_a 3}), which will give coupled second order differential equations.

The exceptional mode $k_V^2 = (n-1) K $, corresponding to $K=1$ and $\ell=1$, receives special consideration. 
In this case the perturbations variable $H_T$ does not exist because  $\mathbb{V}_{ij}$ vanishes, and correspondingly, Eq. (\ref{eq:eq for vector F_a 1}) does not exist. 
$F^a$ is not invariant under the transverse gauge transformation, and it has only one physical degree. Taking $H_T=0$ in (\ref{eq: def vector F_a}), the single physical mode which is invariant under the transverse gauge is given by (\ref{eq:def F^(1)}). 
For the zero mode, the equation for $F^{(1)}$ is $D_a (r^{n+1}F^{(1)})=0$, and its solution is $F^{(1)} = \mathrm{const.} / r^{(n+1)} $.
This solution corresponds to adding a rotation to the background solution, although it is not a dynamical freedom. 
For the massive spectrum of this exceptional mode, the transversely gauge invariant equation is from (\ref{eq:eq for vector F_a 2})
\begin{eqnarray}
 D^c  \left[ \frac{1}{r^{n+2}}D_c \left( r^{n+1} F^{(1) }\right)
         \right]
= - \frac{1}{r}  F^{(1)}_{,zz} ~.
\qquad (K=1, \ell=1).
\label{eq:vector Eq. ell=1}
\end{eqnarray}

\subsection{Scalar-type perturbations}

Scalar perturbations are given by
\begin{eqnarray}
\delta g_{ab} = f_{ab} \mathbb{S},
~~\delta g_{ai} = rf_a \mathbb{S}_i,
~~\delta g_{ij} = 2r^2(H_L \gamma_{ij} \mathbb{S}  + H_T \mathbb{S}_{ij}),
~~\delta g_{\alpha z} = 0, 
\label{sca1}
\end{eqnarray}
where the scalar harmonics $\mathbb{S}$, the associated scalar harmonic vector $\mathbb{S}_i$, and the traceless tensor $\mathbb{S}_{ij}$ are defined by
\begin{eqnarray}
    (\widehat{\Delta}_n+k_S^2) \mathbb{S} = 0,
\quad
    \mathbb{S}_i \equiv -\frac{1}{k_S}\widehat{D}_i \mathbb{S},
\quad
\mathbb{S}_{ij} \equiv \frac{1}{k_S^2} \widehat{D}_i\widehat{D}_j \mathbb{S} + \frac{1}{n} \gamma_{ij} \mathbb{S},
\end{eqnarray}
with the eigenvalues $k_S^2$ given by $k_S^2 = l(l+n-1)$ for $K=1$. By definition, $\mathbb{S}_i$ and $\mathbb{S}_{ij}$ have the following property:
\begin{eqnarray}
  \widehat{D}^i \mathbb{S}_i = k_S \mathbb{S}, 
\quad
  {\mathbb{S}^i}_i=0,
\quad
    \widehat{D}^i \mathbb{S}_{ij} = 
        \frac{n-1}{n} \frac{k_S^2-nK}{k_S} \mathbb{S}_i.  
\end{eqnarray}
We introduce $F(t,r,z)$ and $F_{ab}(t,r,z)$ defined by
\begin{eqnarray}
&& F = H_L + \frac{1}{n}H_T + \frac{1}{r} D^a r X_a, 
\cr
&& F_{ab} = f_{ab} + D_a X_b +  D_b X_a,
\label{eq:def F and F_ab}
\end{eqnarray}
where $ X_a = \frac{r}{k_S}\left(f_a + \frac{r}{k_S}D_a H_T\right)$. 
By using these expansions, we have calculated Einstein equations for the scalar perturbations, which are summarized in the Appendix~\ref{app:Einstein Equations}.

Let us first consider the equations for the generic modes $k_S^2(k_S^2-nK) \neq 0$. The equations directly obtained from the Einstein equations contain such as $f^{ab}_{,zz}$ and $H_{L,zz}$ as we see in Eqs. (\ref{eq:scalar-dG_ab}) and (\ref{eq:delta G_ij trace}). 
Eliminating such terms by utilizing (\ref{eq:delta G_ai}) and (\ref{eq:delta G_ij traceless}), we  obtain the following perturbation equations for $F_{ab}$ and $F$:
\begin{eqnarray}
&& 
D^a D^b F_{ab}  - \square F^c_c   
  - n \frac{D^a r}{r} (D_a F^c_c - 2 D^b F_{ab})
  + \left[   R^{(2)}_{ab} - 2(n+1) \frac{D_a D_b r}{r} 
  + (n^2-3n-2)\frac{D_a r D_b r}{r^2}\right]F^{ab} 
\nonumber
\\
&&
\qquad
    + \frac{k_S^2}{r^2} F^c_c
 + 2 \square F 
  - \left[4k_S^2 - 2(n-1)(n+2)K+4(n+1)(n-2)(Dr)^2\right]\frac{F}{r^2}
   +  2 (n+1)F_{,zz} 
   = 0 ,
\label{eq:D^aD^b F_ab}
\end{eqnarray}
\begin{eqnarray}
& & 
 \square F_{ab}   + \frac{D^c r}{r} (n D_c F_{ab} - 4 D_{(a} F_{b)c}  )
   - 2R^{(2)}_{c(a} F^c_{b)}  
    + 2{R^{(2)}_{acbd}} F^{cd} - \frac{k_S^2}{r^2}F_{ab}    
    + 2(n-2) 
      \left[ \frac{ F_{c(a} D_{b)} D^c r}{r} - \frac{F_{c(a}D_{b)} r D^c r}{r^2}\right]
\nonumber
\\ 
&&
 \qquad 
    + \frac{8}{r} D_{(a} r D_{b)} F
  - 4(n-2)\left(\frac{D_a D_b r}{r} - \frac{D_a r D_b r}{r^2}\right)F 
   + \frac{g_{ab}}{n+1}
\Bigg[
    D^c D^d F_{cd}  - \square  F^c_c  
      - \frac{n}{r}D^c r(D_c F^d_d-2 D^d F_{cd})  
\cr
&&
 \qquad
   + \left( R^{(2)}_{cd} + n(n-1)\frac{D_c r D_d r}{r^2} \right)F^{cd}
+ \frac{k_S^2}{r^2} F^c_c
   - 2n{\square F}-\frac{2n(n+1)}{r}D ^c r D_c F + 2(n-1) \frac{k_S^2-nK}{r^2}F  
 \Bigg]
\cr
&&
 \qquad
    + F_{ab,zz}
   = 0  , 
\label{eq:Box F_ab}  
\end{eqnarray}
where ${(a \, b)}$ is a notation for the totally symmetric parts of tensors \cite{Wald:1984GR}. 
For the zero mode, (\ref{eq:delta G_ai}) and (\ref{eq:delta G_ij traceless}) work as ``constraint'' equations. In the present case, they constitute $\partial_z^2 X_a$, which is given by (\ref{eq:def D_zz Za}).

Additional EOMs are obtained from $\delta R_{Az}=0$. 
From (\ref{eq:R_za-scalar component}) and (\ref{eq:R_zi-scalar component}), we get
\begin{eqnarray}
&& 
\partial_{z}^2 \biggl\{
D^b F_{ab} + n \frac{D^c r}{r} F_{ac} - 2n \frac{D_a r}{r} F
- D^b D_a X_b - \square X_a 
 + n \frac{D^c r}{r} 
    \left( 2 \frac{D_a r}{r} X_c -  D_a X_c - D_c X_a \right)
 +  \frac{k_S^2}{r^2}  X_a
\biggr\}
\cr
&&
\qquad 
+
 k_S^2 r^2   
 D_a\left[ \frac{1}{r^4} \left(  \frac{F_c^c}{2}  + (n-2) F \right)  \right] 
=0
\\
&& F_{,zz} 
 + \frac{2}{r^{n-2}} \frac{D^a r}{r} D_b ( r^{n-2} F_a^b)
 + \frac{ D^a D_b ( r^{n-2} F_a^b) }{2r^{n-2}} 
 - (n-2) \left[ \frac{\square r}{r} + (n+1) \frac{(Dr)^2}{r^2}  \right] F
 - n \frac{D^a r}{r} D_a \left(   \frac{F_c^c}{2} + n F  \right)
\cr
&&
\qquad 
+
\biggl[
    (n-1) \frac{ (D r)^2 }{r^2} + \frac{\square r}{r} + \frac{k_S^2- (n-1)K}{r^2}
\biggr]
\left(  \frac{F_c^c}{2}  + (n-2) F \right)
- \square \left(  \frac{F_c^c}{2}  + (n-1) F \right)
=0
\end{eqnarray}
where (\ref{eq:def D_zz Za}) is used to calculate $X_{a,zz}$.
Finally, $\delta R_{zz}=0$ gives Eq. (\ref{eq:R_zz-scalar component});
\begin{eqnarray}
 &&  \partial_z^2 (f^c_c + 2n H_L ) =0.
\end{eqnarray}
Substituting (\ref{eq:sol H_L,zz}) and (\ref{eq:sol f_a,zz}) into this equation, we obtain an equation which does not contain $z$-derivatives, in contrast to the above four equations. 
These five equations are the basic equations for $\ell \ge 2$ modes. 
We will analyze these in the next section.

For the exceptional mode $k_S^2(k_S^2-nK) = 0$, we need special consideration for the metric perturbations since some harmonic functions vanish in this case. 
For $k_S^2=n K$, which corresponds to $\ell=1$, Eq. (\ref{eq:delta G_ij traceless}) does not exist since $\mathbb{S}_{ij}$ vanishes and 
$H_T$ is not defined.  
For $k_S^2=0$, which corresponds to $\ell=0$, both Eqs. 
(\ref{eq:delta G_ai}) and (\ref{eq:delta G_ij traceless}) do not appear since $H_T$ and $f_a$ do not exist. 
In the following we will consider these two exceptional modes separately.

\subsubsection{$\ell=1$}

For $\ell = 1 $ ($k_S^2=n K$), the metric perturbation 
$H_T$ and hence Eq. (\ref{eq:delta G_ij traceless}) does not exist.
In this case, Eq. (\ref{eq:def F and F_ab}) is replaced by just setting $H_T=0$. 
The transverse gauge transformation of $F$ and $F_{ab}$ becomes
\begin{eqnarray}
 && \delta F  =  - \frac{r}{k_S}
  \left\{
      \frac{k_S^2 }{n r^2} L + D^a r D_a \left( \frac{L}{r} \right) 
  \right\},
\cr
 && \delta F_{ab}  = 
   - D_a \left[ \frac{r^2}{k_S} D_b \left(\frac{L}{r} \right) \right] 
   - D_b \left[ \frac{r^2}{k_S} D_a \left(\frac{L}{r} \right) \right] , 
\label{eq:gauge trans L=1 scalar} 
\end{eqnarray}
and they are no longer transversely gauge invariant.
We will use this gauge degree of freedom when we explicitly solve this mode.

Equations for $F$ and $F_{ab}$ are obtained as follows. 
From Eqs. (\ref{eq:R_zz-scalar component}), (\ref{eq:R_za-scalar component}) and  (\ref{eq:R_zi-scalar component})
\begin{subequations}
\begin{eqnarray}
&& \partial_{z}^2
\left[
 2F - D_c X^c -(n+2) \frac{D^c r}{r}  X_c
\right] = 0 ,
\label{eq:L=1, D_zz eq1}
\\
&& \partial_{z}^2
\Bigl[
  F^c_c + (n-2) D_c X^c + n^2 \frac{D^c r}{r} X_c 
\Bigr] = 0 , 
\label{eq:L=1, D_zz eq2}
\\
&& \partial_{z}^2
\Bigl[
 D^b F_{ab} + n \frac{D^c r}{r} F_{ac} - n^2 \frac{D_a r D^c r}{r^2}X_c
 - n \frac{D_a r}{r} D^b X_b 
\nonumber
\\ 
&& \qquad
 - \Box X_a - D^b D_a X_b + \left(\frac{k_S}{r}\right)^2 X_a
 - n \frac{D^c r}{r} ( D_a X_c + D_c X_a )
\Bigr] = 0 .
\label{eq:L=1, D_zz eq3}
\end{eqnarray}
\label{eq:L=1, D_zz eq123}
\end{subequations}
Here $X_a$ is given by (\ref{eq:delta G_ai}):
\begin{eqnarray}
  X_{a,zz} = 
  - \frac{1}{r^{n-2}}  D_b(r^{n-2}F_a^b) 
  + r D_a \left(\frac{F^c_c}{r}\right) + 2 (n-1)D_a F . 
\label{eq:def X_azz}
\end{eqnarray}
Utilizing this $X_a$, Eqs. (\ref{eq:delta G_ij trace})  (or (\ref{eq:sol H_L,zz})) and (\ref{eq:equation for ell=1}) can be written in terms of $F$ and $F_{ab}$.   
These five equations are the basic equations for $\ell=1$ mode.

\subsubsection{$\ell=0$}

For the $s$-wave ($\ell=0$) perturbation, $H_T$ and $f_a$ do not exist since ${\mathbb S}_i$ and ${\mathbb S}_{ij}$ cannot be defined for this mode. 
Hence $F_{ab}$ and $F$ are given by $F_{ab}=f_{ab}$ and $F=H_L$.
The equations for these variables are given by (\ref{eq:scalar-dG_ab}) and (\ref{eq:delta G_ij trace}).
[or equivalently, (\ref{eq:sol H_L,zz}) and (\ref{eq:equation for ell=1})]. 
Other complementary equations are from $\delta R_{az}=0$ and $\delta R_{zz}=0$, i.e., Eqs. (\ref{eq:R_zz-scalar component}) and (\ref{eq:R_za-scalar component}), 
\begin{eqnarray*}
&& \partial_z^2 (F^c_c + 2n F ) = 0 , 
\cr
&& \partial_z^2 \left(
  D_c F^c_a 
  + n \frac{D^c r}{r} F_{ca}
  - 2n \frac{D_a r}{r} F
 \right) = 0 .
\end{eqnarray*}
The variables $F_{ab}$ and $F$ are gauge dependent, and their four components are reduced to two physical degrees of freedom by fixing transverse gauge (on $z=$const.).
For example, the harmonic gauge condition, $ {\bar{\nabla}}^Ah_{AB}=0$, is a useful gauge fixing, which gives 
\begin{eqnarray}
     D^c   F_{ac} + n \frac{D^c r}{r} F_{ac} - 2n \frac{D^c r}{r} F=0.
\label{eq:divergenceless condition}
\end{eqnarray}
We will use this gauge fixing later.

\section{Stability analysis: Tensor and Vector}
\label{Stability analysis: Tensor and Vector}

\subsection{Background spacetimes and stability condition}

In this section, we discuss about stability of a higher dimensional black string. As a black string solution, we consider a following metric form: 
\begin{eqnarray}
 && g_{ab}dx^a dx^b = - f(r)dt^2 + \frac{dr^2}{f(r)}, 
\label{eq:sch metric}
\\
 && f(r) = K  - \left( \frac{r_h}{r} \right)^{n-1}. 
\nonumber
\end{eqnarray}
Here $g_{ab}$ is the Lorentzian metric of the two-dimensional orbit spacetime, as mentioned in the previous section and the constant parameter $r_h$ defines the horizon radius. Hereafter, we only focus on the $ K =1$ case, because our interest is in the stability of the black string whose intersect is the higher dimensionally Schwarzschild black hole. 
\footnote{
It is interesting to study the stability of the black string in the other backgrounds, for example, with $K\neq 1$ and the cosmological constant.
However, for such cases, even the stability of the Schwarzschild black hole has not been established completely \cite{Kodama:2003kk}.
}

If the equations of perturbations are reduced to a 2nd-order Schr\"{o}dinger-type differential equation, the analysis of the stability can be carried out easily. Writing the Fourier component proportional to $e^{- i \omega t}$ as $\Phi$, let us consider the equation of the following form, 
\begin{eqnarray}
 \omega^2 \Phi 
&=&  \mathcal{A} \, \Phi 
\equiv  
 \left(- \frac{\partial^2}{\partial r_*^2} 
  + V  ( r_*) \right)\Phi, 
\label{eq:Schrodinger-type eq} 
\end{eqnarray}
where the operator $\mathcal{A}$ is the self-adjoint differential operator and  $V(r_*)$ is a smooth function of a coordinate $r_*$. 
(As we see later, $r_*$ corresponds to the tortoise coordinate,  $dr_* = f^{-1} dr $.)
Then, if the operator ${\mathcal{A}}$ with domain $C^\infty_0(r_*)$ is a positive symmetric operator in the $L^2$-Hilbert space with respect to the inner product
\begin{eqnarray}
    (\Phi_1,\Phi_2)_{L^2} 
\equiv
     \int dr_* ~ {\Phi_1}^\dagger (r_*) \Phi_2(r_*),
\label{eq:def inner product}
\end{eqnarray}
the system does not have normalizable negative mode solutions.
Consequently, the amplitude of the solution remains bounded for all times as long as a smooth initial data of compact support in $r_*$ is concerned~\cite{Wald:1979,Wald:1979Erratum}.
(See \cite{Ishibashi:2003ap} for the argument of initial data.)

We should notice that this stability condition of positive self-adjointness is not a necessary condition, but is just a sufficient condition in general. 
In fact, for some type of potential which is not positive definite, it is possible to prove stability of the system by shifting the bottom of potential. 
The method is known as {\it{$S$-deformation}}:
Introducing a new differential operator 
\begin{eqnarray}
    \widehat{D} \equiv \frac{d}{d r_*} + S (r_*)
\end{eqnarray}
with $S$ being some function of $r_*$,  
the inner product is evaluated after integration by parts as 
\begin{eqnarray}
    (\Phi,  {\mathcal A} \Phi)_{L^2} 
    &=&
     \int dr_*  \left( 
         |\widehat{D} \Phi|^2 +  \overline{V}  |\Phi|^2
         \right), 
\cr
 \overline{V} & \equiv& V +  \frac{dS}{dr_*} - S^2  . 
\end{eqnarray}
where the boundary term vanishes for $\Phi \in C^\infty_0 (r_*)$. 
Therefore the $S$-deformation shifts the bottom of potential.

\subsection{Tensor perturbations}

The master equation (\ref{eq:EOM tensor}) for the tensor-type perturbation is the same form of Eq.~(\ref{eq:Schrodinger-type eq}). 
Fourier-expanding along $z$ direction, the operator $\mathcal{A}$ is given by
\begin{eqnarray}
{\mathcal A}
=  - \frac{\partial^2}{\partial r_*^2} 
   + f  k_z^2
   + V_T . 
\label{eq:tensor master eq}
\end{eqnarray}
where $r_* = \int dr f^{-1}(r)$ and $k_z$ is the wave number in $z$ direction which corresponds to the mass spectrum of Kaluza-Klein (KK) modes on $z = z_c$ plane. 
The mass spectrum is taken to be $k_z^2 \ge 0$ without loss of generality. 
Otherwise the linear perturbations break down at some $z$, even at an initial time. 
$k_z^2>0$ is called massive modes, and $k_z^2=0$ is zero-mode which corresponds to the perturbations of the higher dimensional Schwarzschild black holes.

For the background given by (\ref{eq:sch metric}), the potential $V_T$ is expressed as 
\begin{eqnarray}
 V_T(r) = \frac{f}{r^2}
     \left[
        \frac{n(n+2)}{4} f
      +   \frac{n(n+1) }{ (r/r_h)^{n-1}}
      +  k_T^2 - (n-2)
     \right]. 
\end{eqnarray}
Since the spectrum of $k_T^2$ satisfies $k_T^2 - (n-2) = (l-1)(l+n) > 0$, the potential $V_T$ is positive definite in the Schwarzschild wedge.
Therefore we conclude that the black strings are stable with respect to tensor perturbations.

This result is easily understandable.  
The operator (\ref{eq:tensor master eq}) is nothing but the same one as the higher dimensional Schwarzschild black holes, except the presence of KK modes. 
The massive modes increase the stability of perturbations due to its positive contribution. 
This completely fits in with our physical intuition, and it might be anticipated that other type of perturbations are also stable due to the massive spectrum of KK modes. 
However, as we see below, the master variables of vector and scalar perturbations
for the zero mode cannot be used as master variables for massive modes. The massive modes give new physical degrees of freedom for vector and scalar perturbations and
the transversely gauge-invariant equations give coupled second order differential equations. Then the naive expectation like the tensor perturbation does not hold.

\subsection{Vector perturbations}

    \subsubsection{Stability of $\ell = 1 $} 

The equation for $\ell=1$ is given by (\ref{eq:vector Eq. ell=1}). 
By introducing a new variable
$F^{(1)} = r^{-n/2} \Phi $, we can transform the equation into the form of (\ref{eq:Schrodinger-type eq}) with potential
\begin{eqnarray}
V_V^{(1)} = \frac{f}{r^2} 
\left(
   r^2 k_z^2 + \frac{(n+2)}{4 } \Bigl[  2(1-n) +(2+3n) f \Bigr]
\right). 
\label{eq:vector potential V^1}
\end{eqnarray}
However, the form of potential $V_V^{(1)}$ is not positive definite. 
It becomes negative near the horizon for $k_z^2 r_h^2 < (n^2 + n-2)/2$, and the stability for such light modes are not obvious.

The positive definiteness of the symmetric operator $\mathcal{A}$ with the potential (\ref{eq:vector potential V^1}) is shown by the $S$-deformation. 
We find that the following choice 
\begin{eqnarray}
    S = \frac{(n+2)f}{2r}
\label{eq:s-def for vector}    
\end{eqnarray}
gives positive definite potential $\overline{V} = k_z^2 \, f  \ge  0 $.
Therefore, this mode which corresponds to adding a rotation to the background solution is dynamically stable.

    \subsubsection{Stability of $\ell \ge 2$}

Instead of solving Eq. (\ref{eq:eq for vector F_a 3}), which gives coupled differential equations, let us consider Regge-Wheeler gauge by taking 
$ H_T = 0$ on $z=z_c$ surface. 
In this case, the dynamics of $F_a=f_a$ are given by (\ref{eq:eq for vector F_a 2}), which in general gives two coupled differential equations. 
We introduce the following new variables after Fourier-expanding $F_a$ in $z$-direction.
\begin{eqnarray}
  {F}^t   &=& - r^{-n/2} \Psi(t,r)   , 
\\
 {F}^r  &=& r^{1-n/2} \frac{\Phi (t,r)}{ \sqrt{m_V + k_z^2 r^2}} , 
\end{eqnarray}
From (\ref{eq:eq for vector F_a 2}),  $\Psi$ is solved as 
\begin{eqnarray}
    \Psi
    =
 \int^t_{t_*}  \left\{   \frac{  \left[ n \, m_V + (n+2) k_z^2 r^2 \right] \Phi
     + 2r (m_V +k_z^2 r^2) \Phi'
    }{ 2(m_V + k_z^2 r^2)^{3/2}} \right\}  dt 
  + h(r)  , 
\end{eqnarray}
and we find an equation for $\Phi$, 
\begin{eqnarray}
&& \left(  \square - \frac{V_V}{f} \right)  \Phi = 0 , 
\label{eq:master eq for vector generic modes}
\\
&& V_V 
    =
    \frac{f}{r^2}
    \left\{
   \frac{1}{1+ \mathcal{R}^2} 
   \left[ (n-2) + \frac{3f}{1 + \mathcal{R}^2} \right]
     +
   \left[
      ( 1 +  \mathcal{R}^2)  m_V  
     +  \frac{ 1 - 2 (n+2) f}{ (1+ \mathcal{R}^2) }
    \right]
     + \frac{(n+2)}{4} \Bigl[
      f (3n+2) - 2(n-1) 
    \Bigr]
\right\}
\nonumber
\end{eqnarray}
where ${\mathcal{R}}^2 \equiv k_z^2 \, r^2/{m_V}$. 
Here $t_*$ is an initial time and $h(r)$ is an arbitrary function. 
The Eq. (\ref{eq:eq for vector F_a 2}) contains only first time derivative of $F^t$, and hence the initial data of $F^t$ can be specified only by $h(r)$. 
\footnote{
There will be another arbitrary function. 
Substituting the solution of (\ref{eq:master eq for vector generic modes}) into the Eq. (\ref{eq:eq for vector F_a 1}), we can integrate it by $z$ to get $H_T$ at $z \neq z_c$. 
Two arbitrary functions of $x^\alpha$ appears, but one of them can be eliminated by (\ref{eq:residual gauge PQ1}).  
The remaining function corresponds to an ``initial data'' in the bulk, whose evolution is stable sine it is homogeneous (zero mode) in $z$-direction
}

The potential $V_V$ becomes negative near the horizon for $n \gg 1$. 
However, the positive definiteness of this potential can be shown by employing the $S$-deformation. 
Applying the $S$-deformation (\ref{eq:s-def for vector}), the last term in the curly brackets are cancelled out. 
Using the fact that for $K=1$ and $\ell \ge 2$, $m_V$ is bounded below as $m_V \ge n+ 2$, and then the second term is easily shown to be positive definite. Therefore, we conclude that the vector perturbations are stable.

\section{Scalar perturbations}
\label{sec:Scalar perturbations}

\subsection{Gregory-Laflamme mode $\ell= 0 $}
\label{subsec:Gregory-Laflamme mode}

\begin{figure}[t]
\begin{center}
\includegraphics[width=7cm,clip]{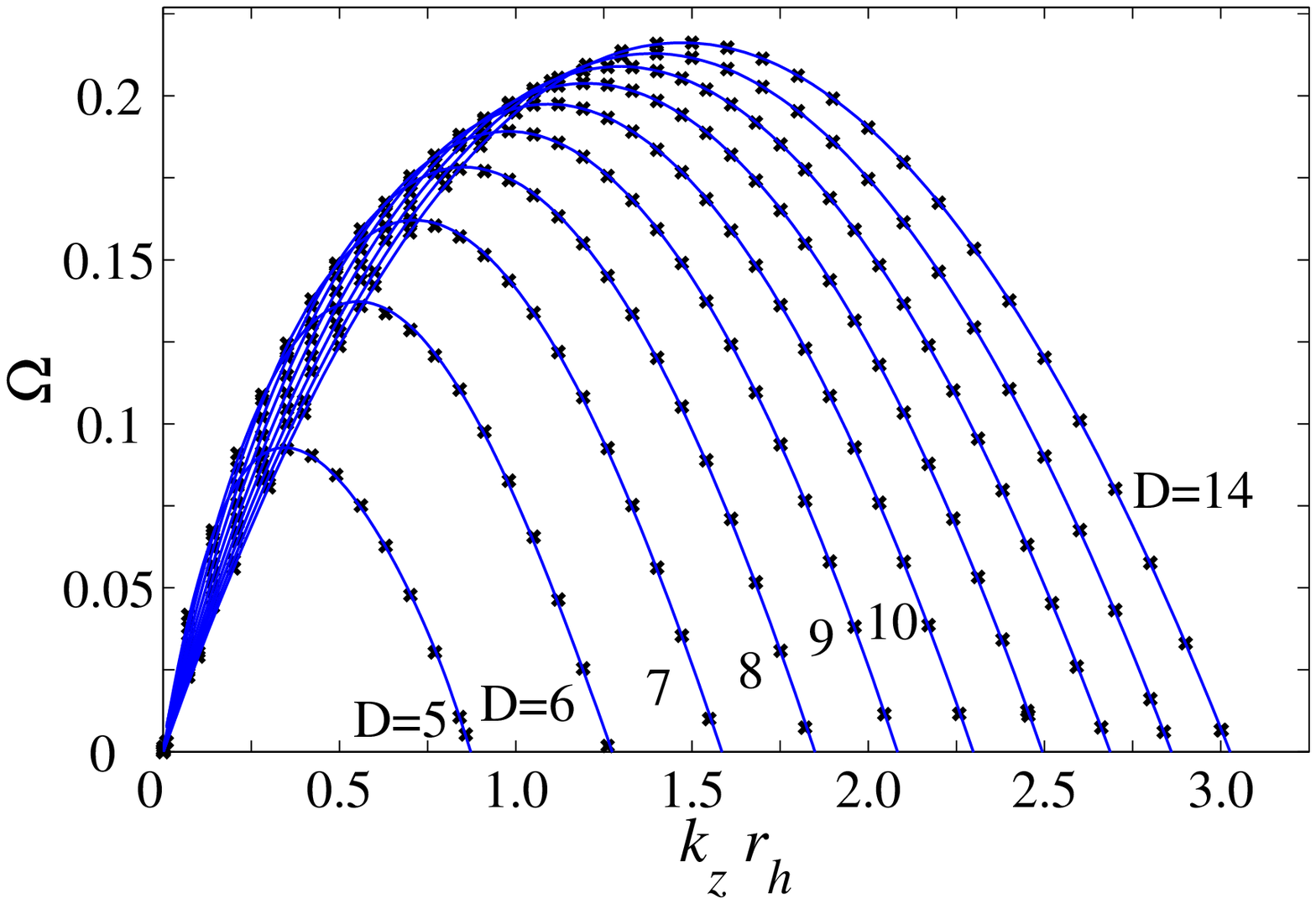}
\caption{
\label{fig:zeromode1}
Plot of $\Omega$ as a function of $k_z$ for black strings with spacetime dimensions $D=n+3=5,~6, \cdots, 14$. 
The wave number $k_z$ is normalized by the horizon radius $r_h$.
}
\end{center}
\begin{center}
\includegraphics[width=7cm,clip]{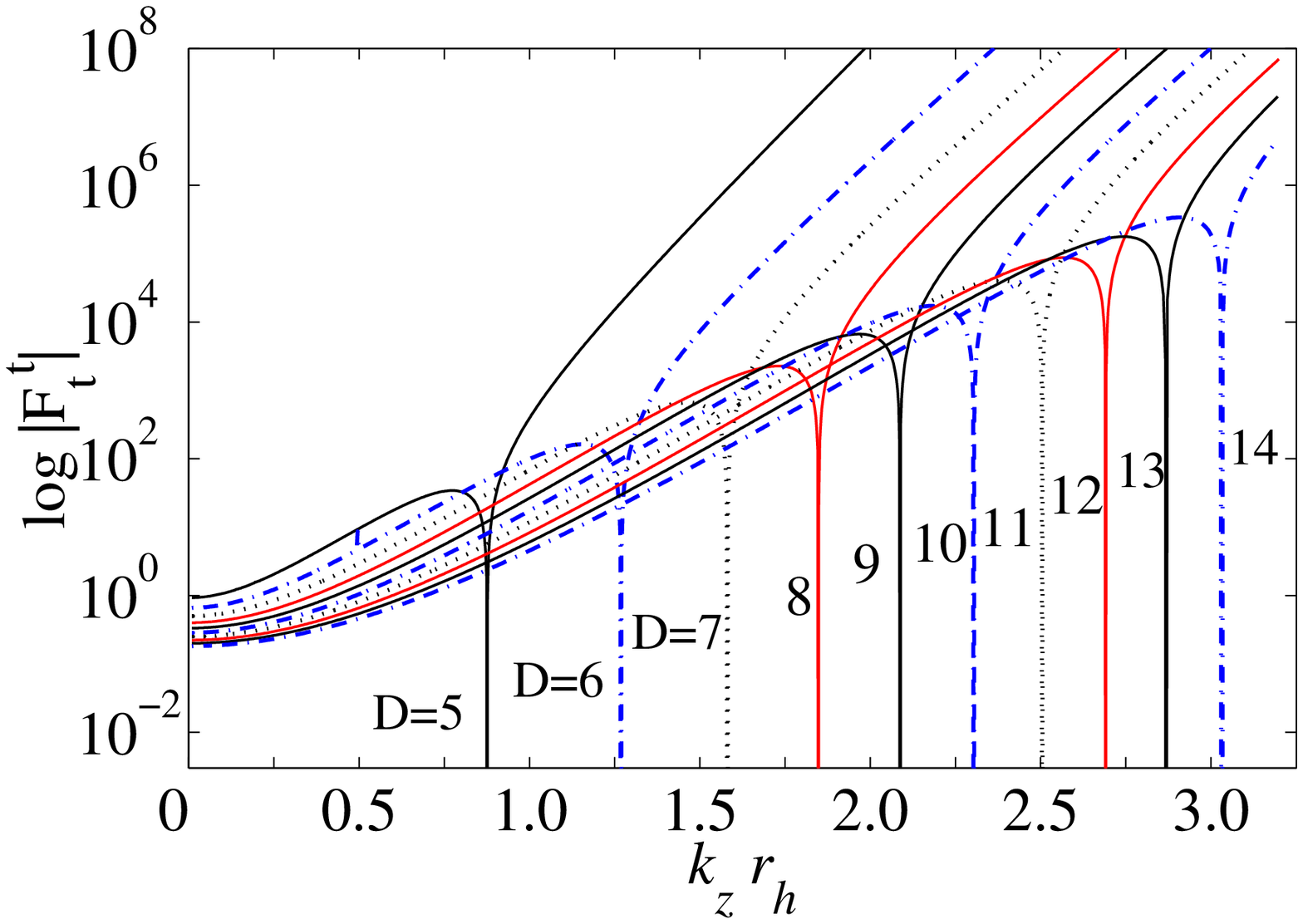}
\caption{
\label{fig:zeromode2}
Static mode search. 
The possible asymptotic solutions of ${F^t_t}$ are ${F^t_t} \propto e^{\pm k_z r}$. In the figure, we plot ${F^t_t}$ at some $r/r_h \gg 1$ with respect to the single shooting parameter $k_z$. 
Since the normalizable solution decays exponentially, each narrow ``throat" corresponds to a normalizable mode. 
The critical wave numbers agree precisely with the static limit $\Omega=0$ in Fig.~\ref{fig:zeromode1}. 
}
\end{center}
\end{figure}

The $s$-wave ($\ell=0$) perturbation is the unstable mode studied by GL. Here, we discuss this mode in our framework and recover their result. 
We can use the residual gauge degrees of freedom (\ref{eq:scalar gauge T_a, L}) to fix unphysical gauge modes. After eliminating the terms proportional to $z$-derivatives of $F_t^t$ and $F$ by using Einstein's equations, we can apply the harmonic gauge condition (\ref{eq:divergenceless condition}) to rewrite  $F_t^t$ and $F$ on $z=z_c$ in terms of $F_{t}^r$ and $F_r^r$.  
Then we finally obtains a second order ordinary differential equation (ODE) of $F_r^r$ (or $F_{t}^r$) in Fourier space, assuming $F_r^r \propto e^{\Omega t + i k_z z }$.
Although it is a second order ODE with respect to $r$, the equation in the original space contains higher derivatives of $t$ and $z$. (See \cite{Hovdebo:2006jy} for more tractable equation.)
From the master equation, the boundary conditions required for a normalizable mode are 
\begin{eqnarray}
&& F^r_r \propto e^{-r \sqrt{k_z^2 + \Omega^2}},  \quad (r \to \infty)
\cr
&& F^r_r \propto \frac{1}{(r-r_h)^{1 -  \Omega/(n-1) }}. \quad  (r \to r_h)
\label{eq:BC for GL mode}
\end{eqnarray}
This is a one-parameter shooting problem with shooting parameter $\Omega >0$. 
We have solved this problem numerically, searching for the growth rate $\Omega$ for given $k_z$. 
The result is shown in Fig.~\ref{fig:zeromode1}, which agrees with the original analysis~\cite{Gregory:1993vy}.
\footnote{
The harmonic gauge does not fix the gauge $T_a$ completely. 
Besides the static radial gauge transformation, the residual gauge is $T_t \propto r^{1-n}$. 
$F_t^r$ depends on this gauge while  $F_r^r$ is free from this mode. 
}

Another type of simple master equation can be obtained by taking static limit. 
To obtain the static mode, we adopt a gauge fixing 
\begin{eqnarray}
    F_{tr} = 0, \quad (z=z_c)
\end{eqnarray}
without fixing the pure radial gauge $T_r(r)$.
In Fourier space, we find a master equation
\begin{eqnarray}
&& 
\left[
 \frac{d^2}{dr^2} + \frac{1}{\mathcal{R}}
\left(
\mathcal{P} r \frac{d}{dr} +  \mathcal{Q}
\right) 
\right]
{F}^t_t
  = 0,
\nonumber
\\
&& 
\quad
\mathcal{P} =
   2 N^3
 + N(N^2 +7N +12 ) f^3
 - 2 ( 3N^2 + 6N -2k_z^2 r^2 ) f^2
 - N ( 3N^2 + N - 4 k_z^2 r^2 ) f, 
\cr
&&
\quad
\mathcal{Q} =  
N 
\Bigl( 
    N^2  (2 N^2 + 6N -3k_z^2 r^2) 
  + N  (2N^3+ 6N^2 - 3k_z^2 r^2 N -8k_z^2 r^2) f^2
\cr
&&
\quad
\qquad
  - 2f(2 N^4 + 6 N^3 - 3k_z^2 r^2 N^2-4 k_z^2 r^2 N+2 k_z^4 r^4 ) 
\Bigr),
\cr
&&
\quad
\mathcal{R} =
r^2  f \left[
   N^2+ N (N+4) f^2 - 2(N^2 + 2N - 2 k_z^2 r^2 ) f 
   \right],
\label{eq:static mode eq for L=0}
\end{eqnarray}
where $N=n-1$. Other components $F_r^r$ and $F$ are given in terms of $F_t^t$. 
(Note that another type of master equation was derived in \cite{Kudoh:2005hf}, which is more tractable than the above equation in practice.)
A Neumann condition on the horizon is obtained by requiring the regularity on the horizon. 
Solving this equation is the one-parameter shooting problem with the shooting parameter $k_z$.  
Hence we can think of this equation as a master equation determining the GL static mode. As is known well, the wave number of this static mode, which will be denoted $k_{\mathrm{crit}}$, gives a critical point at which stability of the $s$-wave perturbation changes. 
For $k_z < k_{\mathrm{crit}}$, the perturbations are unstable, whereas they becomes stable for $k_z > k_{\mathrm{crit}}$ (Fig.~\ref{fig:zeromode1}).

The equation (\ref{eq:static mode eq for L=0}) has two asymptotic solutions behaving ${{F}^t_t} \propto e^{ \pm k r}$, and only the decaying mode is the physical normalizable solution.  Such physical solution can be easily found by searching a minimum value of ${{F}^t_t}$ at some fixed asymptotic point as a function of $k_z$. 
Figure~\ref{fig:zeromode2} shows the result of the shooting problem. 
As we see, the critical wave number $k_{\mathrm{crit}}$ can be precisely determined by this method, and this agrees completely with the analysis of dynamical perturbations discussed above.

    \subsection{ $\ell = 1$ }

For the zero mode $k_z^2 = 0$, the $\ell=1$ mode has no physical degrees of freedom. This can be easily observed from the fact that the master variable of the massless mode can be reintroduced by recovering the lacked equation (\ref{eq:delta G_ij traceless}) as a gauge condition.
However, since such gauge fixing is not complete, there remain additional residual gauge degrees of freedom. By using the residual gauge degrees of freedom, it is shown that there is no dynamical degrees of freedom in the vacuum case~\cite{Kodama:2003jz,Kodama:2003kk}.  
(More direct counting of physical degrees of freedom is possible by taking $F=0$ gauge fixing.)

For the KK modes, there is no unstable dynamical mode.  
This is confirmed directory by solving the EOMs on $z=z_c$.  
Let us take the gauge $F=0$. This is not a complete gauge fixing, but $F_{ab}$ does not depend on the residual gauge. 
After eliminating all terms proportional to $\partial_z^2 F$ by employing (\ref{eq:L=1, D_zz eq1}), we can solve the EOMs explicitly after tedious calculations. 
One finds that only trivial solutions are allowed on $z=z_c$ in the present case, so that there is no unstable dynamical degree for the KK modes. 
\footnote{ 
Equation (\ref{eq:L=1, D_zz eq3}) for the KK modes with (\ref{eq:def X_azz}) corresponds to the constraint equation (\ref{eq:delta G_ai}) for the zero mode. 
Eq. (\ref{eq:L=1, D_zz eq1}) is used to take the gauge $F=0$, and Eq. (\ref{eq:L=1, D_zz eq2}) works as a constraint equation.
}

    \subsection{ $\ell \ge 2$ }

For the generic modes of scalar perturbations, the Einstein equations consist of five equations, and they give coupled partial differential equations. 
We decompose $F$ and $F_a^b$ as follows;  
\begin{eqnarray}
F &=& \frac{1}{8 r^{n-2}} \left(  \psi -  5 p- q  \right)
\nonumber
\\
F_t^t &=& \frac{1}{4r^{n-2}} \left(  \psi  -  p - 5 q   \right)
\nonumber
\\
F_r^r &=& \frac{1}{4r^{n-2}} \left(   \psi  +  3p + 3 q  \right)
\nonumber
\\
F^r_t &=& \frac{- \partial_t Z}{r^{n-2}}, 
\end{eqnarray}
Then from the transversely gauge-invariant equations (\ref{eq:D^aD^b F_ab}) and (\ref{eq:Box F_ab}), we obtain  
\begin{eqnarray}
&&
\Box Z  + \partial_{z}^2 Z
- \frac{k_S^2}{r^2} Z
+ \frac{( 1-n + (n+3) f)}{ r} p
+ \frac{2( 1-n +n f)  }{ r} q
+ \frac{(n f -2 n+2)   }{r} Z'   =0, 
\label{eq:eqs for phiA,B,z}
\\
&&  \Box p + \partial_{z}^2 p
    - \frac{ 4 }{3 r f} \ddot{Z}
 -  \left[ { k_S^2 } + \frac{ 4 - 4n + (3n-10) f }{ 3 } \right] \frac{p}{r^2}
  + \Bigl[  2(1-n) +(3n-2)f  \Bigr]  \frac{2q}{3r^2}
  - \frac{f}{3r} \Bigl[  (3n-4) p' + 4q'\Bigr] 
= 0 ,
\cr
&&
 \Box q  + \partial_{z}^2 q 
 + \frac{ 2 [ 3(1-n) +  f(1+3n) ]}{ 3 r f^2} {\ddot Z}
 - \frac{f}{3r} \Bigl[ 8p' + (3n-8) q' \Bigr]
\cr
&& \qquad 
+   \left[  \frac{(3n^2 + 6n -25)f}{6} - \frac{2(n-1)}{3} - \frac{(n-1)^2}{2f} \right]  \frac{p}{r^2} 
+   \left[  \frac{(3n^2 - 9n +5)f}{3} + \frac{2(n-1)}{3} - \frac{(n-1)^2}{f} 
         - k_S^2  
    \right]  \frac{q}{r^2} 
   = 0 ,
\nonumber
\end{eqnarray}
and $\psi$ is given by 
\begin{eqnarray}
 \frac{(n+1)}{2}
  \partial_z^2  \psi &= & 
    \frac{(1+5n)}{2}  \partial_z^2 p
  + \frac{(5+n)}{ 2}   \partial_z^2 q   
 - (n-1 + (1+3n) f )  \frac{p'}{r}
 - 2(n-1 + f ) \frac{q' }{r}
\cr
&&  
 + \frac{ p}{ r^2}  
    \left[ (n^2 + 3n-4) -(n-5)f - \frac{(n-1)^2}{f}   \right]  
 + \frac{ 2q}{ r^2}  
    \left[ (n^2 - 3n+2) +(2n-1)f - \frac{(n-1)^2}{f}   \right]  
\cr
&&  
 - \frac{2k_S^2}{r^2}   \Bigl[  2p +q  \Bigr] 
 - \frac{2}{f}  \partial_t^2
    \Bigl[  p+2q  \Bigr] 
 - \frac{ 4 }{ f^{3/2} }  \partial_t^2  \partial_r
    \Bigl[ \sqrt{f} Z \Bigr] .
\label{eq:D_zz psi}
\end{eqnarray}
From other remaining equations, we obtain a non-trivial equation for $\psi$.
\begin{eqnarray}
&&  \Box \psi
   + \partial_{z}^2 \psi
+ \frac{2  \left[ 1-n + (n-3) f \right] }{rf^2} \ddot{Z}
+ \frac{\psi}{r^2}
\left[
   (n-2)^2 f + (n-2)(1-n) - k_S^2 
\right] 
+ \frac{(4-n) f }{r} \psi' 
- \frac{8f}{r} ( 2p' + q')
\cr
&&
- \Bigl[
      ( 9n^2 -56n +71)f -2(n-1)(5n-16)+ \frac{(n-1)^2}{f}
  \Bigr] \frac{p}{ 2r^2}
+ \Bigl[
      (11n-17)f + (n-8)(n-1) - \frac{(n-1)^2}{f}  
  \Bigr] \frac{q}{ r^2}
= 0
\nonumber
\end{eqnarray}

\begin{figure}[tb]
\begin{center}
\includegraphics[width=7cm,clip]{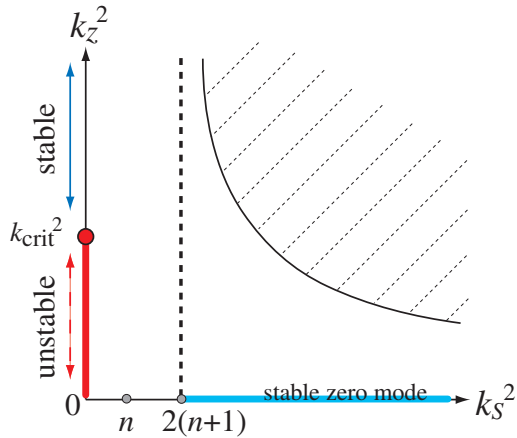}
\caption{
\label{fig:stable-unstable_phase}
Stable-unstable phase on $k_z^2$-$k_S^2$ plane. 
The zero mode $k_z^2=0$ corresponds to the perturbations of higher dimensional Schwarzschild BHs, which are stable. 
The Gregory-Laflamme mode $(\ell=0)$ is at $k_S^2=0$ with $k_z^2 < k_{\mathrm{crit}}^2$. 
On the plane, the shaded upper-right corner with $k_S \gg 1$ or $k_z^2 \gg r_h^{-2}$ is shown to be stable analytically.
The stability of other generic modes is confirmed numerically.  
}
\end{center}
\end{figure}

We first notice that Eq. (\ref{eq:eqs for phiA,B,z}) does not contain $\psi$, and  $\psi$ can be determined by (\ref{eq:D_zz psi}) once we solve $p$, $q$ and $Z$. 
Hence it is sufficient to analyze Eq. (\ref{eq:eqs for phiA,B,z}) for the stability problem. 
We begin with a limited case to study the stability. 
In the limit $k_S \gg 1$, the EOMs are 
\begin{eqnarray}
\left[ \square - \left( k_z^2 + \frac{k_S^2}{r^2} \right) \right] Z &=& 0, 
\\
\left[ \square - \left( k_z^2 + \frac{k_S^2}{r^2} \right) \right] p &=&
\frac{ 4 }{ 3r f} \ddot{Z} ,
\\
\left[ \square - \left( k_z^2 + \frac{k_S^2}{r^2} + \frac{n^2}{r^2f} \right) \right] {q} 
&=&
  \frac{(n-1)^2}{2r^2 f} ~ p 
+ \frac{ 2 [ 3 (n-1) -(3n+1) f ]}{ 3r f^2} \ddot{Z} , 
\end{eqnarray}
where we have left the terms proportional to $1/f$ since it becomes dominant near the horizon. 
Apparently, $Z$ are stable due to the positive definite potential. 
Then, the stability of $p$ and $q$ is also obvious. 
Furthermore, we notice that the same argument holds true for very massive modes $k_z^2 \gg r_h^{-2}$, without taking the limit $k_S \gg 1$. 
Thus the system is stable if  $k_S \gg 1$ or $k_z^2 \gg r_h^{-2}$. 
We note that the system is stable in the zero mode limit $k_z^2 \to 0$, since in this limit the perturbations are the same as the Schwarzschild black holes.  
Hence on the $k_z^2$ - $k_S^2$ plane, there exists stable region. 
The stable/unstable parameter region discussed here is summarized in Figure~\ref{fig:stable-unstable_phase}.

\begin{figure}[tb] 
\begin{center}   
\includegraphics[width=7cm,clip]{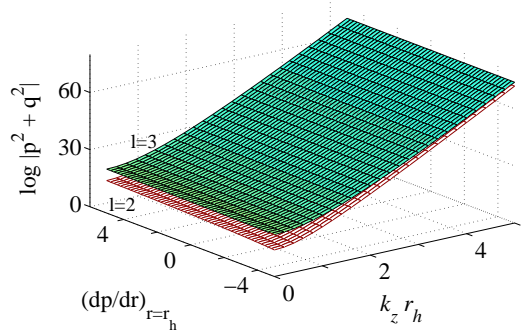}
\caption{ 
\label{fig:search for static mode} 
Search for critical static mode for $\ell = 2, 3$. 
(See Fig.~\ref{fig:zeromode2} for $\ell=1$ mode.)
This is a two-parameter shooting problem. The two parameters are the wave number $k_z$ and the derivative of $p$ at the horizon. 
The figure shows a plot of $(p^2 + q^2)$ at some asymptotic region with respect to the two parameters. 
Possible asymptotic solutions are $p,q \propto e^{\pm k_z r}$, and normalizable solutions will decay at $r \gg r_h$. 
No narrow ``throat'' appears so that there is no normalizable static mode.
For other higher multipoles ($\ell \ge 2$), we obtained the same results. 
}
\end{center}
\end{figure}

For general modes with arbitrary $k_z^2$ and $k_S^2$, we performed a numerical search for unstable solution, as we do in Sec.~\ref{subsec:Gregory-Laflamme mode}, 
Assuming an unstable perturbation $\propto e^{\Omega t}e^{i k_z z} $, we obtain boundary conditions similar to (\ref{eq:BC for GL mode}). Then we performed a parameter search in the relevant region of $(k_z, \Omega)$ and no solutions were found, suggesting that no instability exists for the generic modes. 
To confirm this result furthermore, we have also performed a search for critical static mode: if the system is unstable, a static mode will exist since the real eigenvalue in the stable region will cross the zero axis at lease once when it becomes unstable.  
Since the horizon boundary conditions are not the same as those obtained by just taking the static limit of dynamical perturbations, this numerical search works as an independent search of unstable mode. 
Redefining $\partial_t Z = \zeta$, we take static limit of (\ref{eq:eqs for phiA,B,z}). In this limit $\{p, q \}$ and $\zeta$ are decoupled, and we can easily performed the search. 
The differential equations for $\{p, q\}$ are a two-parameter shooting problem, and a part of the result is shown in Fig.~\ref{fig:search for static mode}, which corresponds to Fig.~\ref{fig:zeromode2}.
Clearly, there is no static solution satisfying appropriate boundary conditions. 
The same result holds also for $\zeta$. 
Therefore we conclude that the black strings are stable for all types of perturbations except the $s$-wave mode.

\section{Summary and discussion } 
\label{sec:Summary}

In summary, we have studied stability of black strings with respect to all types of gravitational perturbations.
There are three types of perturbations; tensor, vector, and scalar perturbations. 
The vector and scalar perturbations have the exceptional modes of multipole moment besides the generic modes. For the higher dimensional Schwarzschild black holes, the exceptional modes are not dynamical degrees of freedom.
However, we have paid particular attention to the exceptional modes since they might become dynamical with some instability.

The generic modes of tensor ($\ell \ge 1$) and vector ($\ell \ge 2$) perturbations have been shown to be stable. 
The generic modes of scalar ($\ell \ge 2$) perturbations were studied partially employing numerical investigation, and they have been shown to be stable. 
For the exceptional modes, we have discussed that the vector perturbation of $\ell=1$, which corresponds to adding a rotation, is stable, and the exceptional mode $\ell=1$ of scalar perturbation has no unstable dynamical degree of freedom.  
The $\ell=0$ mode of scalar perturbation is also the exceptional mode, and it is dynamically unstable as discussed by Gregory and Laflamme. 
After all, the unstable mode of gravitational perturbations for black strings is only the $\ell=0$ mode of scalar perturbation.

The zero mode ($k_z^2=0$) of the scalar perturbation with $\ell=0$ corresponds to a shift of the mass parameter of the higher dimensional Schwarzschild black holes (or uniform black strings), and hence this mode is not allowed as a consequence of the Birkhoff's theorem.  However, the KK mode with $\ell=0$ is essentially different from the gravitational perturbations of the Schwarzschild black holes, and in fact it does not change the mass of the black strings.
Therefore, from the viewpoint of effective theory on a $z=\mathrm{const.}$  plane, we understand that {\it{the existence of Gregory-Laflamme instability is directly related to the inapplicability of Birkhoff's theorem.}}

This observation is useful to consider a possible counterexample of correlated-stability conjecture (CSC). 
If we do not interpret CSC in strong sense, the instability predicted by CSC is $s$-wave instability~\cite{Reall:2001ag}.  
To have some insight, let us discuss a black hole obtained by dimensional reduction
of the black string/brane. 
If the black hole is a hairy black hole and (generalized) Birkhoff's theorem cannot be applied, the $s$-wave perturbation becomes a dynamical degree of freedom. 
This $s$-wave perturbation is homogeneous (zero-mode) perturbation in the original spacetimes, and it is not the (massive) perturbation for which CSC concerns.
Then if there is a model in which the homogeneous $s$-wave becomes unstable for some parameter region, the model will be a counterexample of CSC since the instability is disconnected from CSC.
In fact, the recently proposed counterexample is based on a hairy black hole~\cite{Friess:2005zp,Gubser:2005ih}, and the unstable mechanism is along the line of the above discussion.

We finally address possible extension of present analysis.
First, it is interesting problem to study the stability of
charged black strings, focusing on how a given charge
works to make the string stable near the thermodynamically
stable and/or BPS state. It will give us deeper understanding
of CSC from the perspective of dynamics.
Second, we have analyzed the black object with a single
trivial transverse direction, for simplicity. For black branes
with translationally invariant multiple directions, it will be
possible to expand the perturbation variables by harmonic
tensors associated with the uniform transverse directions.
We would like to discuss these issues somewhere else.

\acknowledgments
It is a pleasure to thank Kenta Kiuchi for helpful correspondence and discussion at an early stage of this project. 
The author would like to thank Shinji Mukohyama, Hideo Kodama, Akihiro Ishibashi, Umpei Miyamoto and Robert C. Myers for their very helpful comments, discussion and conversation. 
In particular, Akihiro Ishibashi presented a useful observation. 
He would like to thank the organizers of the ``Scanning New Horizons: GR Beyond 4 Dimensions'' workshop in Santa Barbara for a stimulating environment, a relaxed atmosphere and hospitality while this work was being completed. 
The author also wants to thank Barak Kol, Dan Gorbonos and Hebrew University of Jerusalem for their hospitality and stimulating discussion during an initial stage of this work. 
This work was supported in part by JSPS (Japan Society for the Promotion of Science) fellowship and the National Science Foundation under Grant No. PHY99-0794

\appendix

    \section{Gauge transformation}

In this appendix we summarize the transverse gauge transformation (\ref{eq:residual gauge PQ2}). 
The metric perturbation $h_{AB}$ transform as 
\begin{eqnarray}
    \delta h _{AB} = - \nabla_A \xi_B  - \nabla_B \xi_A 
\end{eqnarray}
in terms of the infinitesimal gauge transformation $\delta x^A = \xi^A$. 
The transverse gauge transformation (\ref{eq:residual gauge PQ2}) can be decomposed into 
\begin{eqnarray}
    \delta h_{ab} &=& - D_a \xi_b - D_b \xi_a
\\
    \delta h_{ai} &=& - r^2 D_a \left( \frac{\xi_i}{r^2} \right) 
                    - \widehat{D}_i \xi_a
\\
    \delta h_{ij} &=& - \widehat{D}_i \xi_j - \widehat{D}_j \xi_i 
                      - 2r \, \gamma_{ij} \, \xi_a D^a r  . 
\end{eqnarray}
Since the infinitesimal transformation $\xi$ has no tensor component, the expansion coefficient of the tensor perturbation is gauge invariant. Our interest is therefore gauge transformation of vector and scalar perturbations.

The vector component of the transverse gauge transformation is 
\begin{eqnarray}
    \xi_a =0,  \quad \xi_i = r L \mathbb{V}_{i}
\end{eqnarray}
for the modes $k_V^2 \neq (n-1) K $,  where $L=L(x_a)$ is an arbitrary function. 
Then the corresponding expansion coefficients of the perturbation transform as 
\begin{eqnarray}
 \delta f_a = - r D_a \left( \frac{L}{r} \right), 
\quad
 \delta H_T = \frac{k_V}{r} L. 
\end{eqnarray}

As for the scalar perturbation, the gauge transformation for $k_S^2 (k_S^2-n K) \neq 0 $ are given by
\begin{eqnarray}
    \xi_a = T_a \mathbb{S}, \quad   
    \xi_i = r L  \mathbb{S}_i.
\label{eq:scalar gauge T_a, L}
\end{eqnarray}
Under these transformations, the expansion coefficients of the metric perturbation transform as 
\begin{eqnarray}
 \delta f_{ab} &=& -D_a T_b -D_b T_a, 
\cr
    \delta f_{a} &=& - r D_a\left( \frac{L}{r} \right) + \frac{k_S}{r} T_a, 
\cr
    \delta H_L &=& - \frac{k_S}{nr} L - \frac{D^a r}{r} T_a, 
\cr
    \delta H_T &=&  \frac{k_S}{r} L .
\label{eq:gauge trans for scalar pert}
\end{eqnarray}
The gauge transformation for $k_S^2 (k_S^2-n K) =0 $ are obtained by setting appropriate functions equal to zero in the above equations.

\section{Details of calculations}

In this Appendix, we summarize the details of calculating perturbed Einstein's equations for completeness.  Some of them are based on Ref.~\cite{Kodama:2000fa}.

\subsection{Background Quantities}

We consider perturbations of spacetime on $(n+2+1)$-dimensional spacetime whose unperturbed background geometry is given by the metric (\ref{eq:ds^2}).
Decomposition of connection coefficients is
\begin{eqnarray}
{\bar \Gamma}^a_{bc}={}^{(2)}\!\Gamma^a_{bc}(y), ~
{\bar \Gamma}^a_{ij}=-r (D^a r) \gamma_{ij}  ,~
{\bar \Gamma}^i_{aj}=\frac{D_a r}{r}\delta^i_j,~
{\bar \Gamma}^i_{jk}=\hat\Gamma^i_{jk}(x).
\end{eqnarray}
Here ${}^{(2)}\!\Gamma^a_{bc}$ is the Christoffel symbol of the two-dimensional orbit spacetime. 
Curvature and Ricci tensors are 
\begin{eqnarray}
&& 
 {\bar R}^a{}_{bcd} ={}^{(2)} \! R^a{}_{bcd},  \qquad \qquad
 {\bar R}^i{}_{ajb} =-\frac{D_aD_b r}{r}g^i_j, \qquad \qquad
 {\bar R}^i{}_{jkl} =[K-(Dr)^2](g^i_k \gamma_{jl}-g^i_l \gamma_{jk}).
\nonumber
\\
&& {\bar R}_{ab} = {}^{(2)} \!R_{ab}-\frac{n}{r}D_aD_b r, 
\quad
{\bar R}^i_j  =\left[-\frac{\square r}{r}+(n-1)\frac{K-(Dr)^2}{r^2}\right]g^i_j,
\qquad
{\bar R}_{ai} =0,
\nonumber
\\
&& {\bar R} ={}^{(2)}\!R - 2n\frac{\square 
r}{r}+n(n-1)\frac{K-(Dr)^2}{r^2}. 
\end{eqnarray}
Einstein tensors are decomposed as 
\begin{eqnarray}
&& {\bar G}_{ab}=\,{}^{(2)} \!G_{ab}-\frac{n}{r}D_aD_b r 
-\left[\frac{n(n-1)}{2}\frac{K-(Dr)^2}{r^2}
-\frac{n}{r}\square r\right]g_{ab} 
\nonumber
\\
&& {\bar G}^i_j
=\left[-\frac{1}{2}{}^{(2)} \!R-\frac{(n-1)(n-2)}{2}\frac{K-(Dr)^2}
{r^2}+\frac{n-1}{r}\square r \right]g^i_j
\\
&& {\bar G}_{ai}=0.
\nonumber
\end{eqnarray}

For the two-dimensional metric
\begin{eqnarray}
 ds^2 = -f(r) dt^2 + \frac{1}{f(r)} dr^2, 
\end{eqnarray}
Ricci tensor and Riemann tensor are explicitly given by
\begin{eqnarray}
 {}^{(2)}\!R  = - f'' , 
\quad
 R_{a}^b = \delta^b_a \frac{ {}^{(2)}\!R }{2}, 
\quad
R_{abcd} = (g_{ac} g_{bd} - g_{ad} g_{bc})
            \frac{ {}^{(2)}\!R }{2}, 
\end{eqnarray}
and non-vanishing Christoffel symbols are
\begin{eqnarray}
\Gamma_{tr}^t = \frac{f'}{2f} , \quad
\Gamma_{tt}^r = \frac{f f'}{2} , \quad
\Gamma_{rr}^r = - \frac{f'}{2f} .
\end{eqnarray}

\subsection{Perturbations of the Ricci Tensors}
\label{Appendix:Perturbations of the Ricci Tensors}

We consider metric perturbations under the gauge fixing of Eq. (\ref{eq:dg_ab}).
In general the perturbation of the Ricci tensor is expressed in terms of $h_{MN}=\delta {\bar g}_{MN}$ as 
\begin{eqnarray}
 2\delta {\bar R}_{MN} &= & -{\bar\nabla}^L{\bar\nabla}_L 
  h_{MN}-{\bar\nabla}_M{\bar \nabla}_N h
  +{\bar\nabla}_M{\bar \nabla}_L h^L_N 
  + \bar\nabla_N\bar\nabla_L h^L_M 
\nonumber
\\
&& +{\bar R}_{ML}h^L_N+ {\bar R}_{NL}h^L_M-2 {\bar 
   R}_{MLNS}h^{LS},
\nonumber
\\
\delta {\bar R} &=& -h_{MN}{\bar R}^{MN}+\bar\nabla^M\bar\nabla^N 
h_{MN}-\bar\nabla^M\bar\nabla_M h.
\end{eqnarray}
Here and hereafter the trace $h$ is
$     h^A_A  = h_{MN}g^{MN} 
            = h_a^a + r^2  h^{ij} \gamma_{ij}$.

\subsubsection{Decomposition formula}

To calculate the perturbed Ricci tensor, we need to decompose the connection $\nabla$ into $D$ and $\hat D$.
The operator $D$ and $\hat{D}$ work as 
\begin{eqnarray}
&&   \hat{D}_j h_{ab} := \partial_j h_{ab} , 
\cr
&&   \hat{D}_{j}h_{ai} := \partial_j  h_{ai} - \hat{\Gamma}^k_{ji} h_{ak} ,
\cr
&&   D_a h_{ij} : = \partial_{a} h_{ij} ,
\cr
&&   D_a h_{bj} : = \partial_{a} h_{bj} - {}^{(2)}\Gamma^e_{ab}h_{ej} .
\end{eqnarray}
The followings are useful formulas of decomposing the operator $D$ for arbitrary tensor $h_{AB}$ and vector $T_A$.
\begin{eqnarray}
&&    \bar{\nabla}_{a} T_{b}  = D_a T_{b} , 
\nonumber \\
&&    \bar{\nabla}_{i} T_{j}  = \hat{D}_i T_j + r (D^a r) \gamma_{ij} T_a  ,
\nonumber \\
&&    \bar{\nabla}_{i} T_{a}  = \hat{D}_i T_a -  \frac{D_a r }{r} T_i , 
\nonumber \\
&&    \bar{\nabla}_{a} T_{i}  = \hat{D}_a T_i -  \frac{D_a r }{r} T_i , 
\nonumber \\
&&
    \bar{\nabla}_z T_A = \partial_z T_A
\end{eqnarray}
and 
\begin{eqnarray}
&&    \bar{\nabla}_{a} h_{cd}  = D_a  h_{cd} ,
\nonumber \\
&&    \bar{\nabla}_{a}\bar{\nabla}_{b} h_{cd}  = D_a D_b h_{cd} ,
\nonumber \\
&&    \bar{\nabla}_{a} h_{ij}  = D_a h_{ij} - 2 \frac{D_a r}{r} h_{ij} ,
\nonumber \\
&&    \bar{\nabla}_{a} h_{bj}  = D_a  h_{bj} - \frac{D_a r}{r} h_{bj} ,
\nonumber \\
&&    \bar{\nabla}_{i} h_{bc} =  \hat{D}_{i}h_{bc} 
                                - \frac{D_b r}{r} h_{ic}
                                - \frac{D_c r}{r} h_{bi} ,
\nonumber \\
&&    \bar{\nabla}_{i} h_{jc} =  \hat{D}_{i}h_{jc} + r(D^a r) \gamma_{ij} h_{ac}
                                - \frac{D_c r}{r} h_{ij} ,
\nonumber \\
&&    \bar{\nabla}_{i} h_{jk} =  \hat{D}_{i}h_{jk} + r (D^c r) \gamma_{ij} h_{ck}
                                + r(D^c r) \gamma_{ik} h_{jc} ,
\nonumber \\
&&    \bar{\nabla}_{i}\bar{\nabla}_{j} h
     =  \hat{D}_{i} \hat{D}_{j} h 
        + r (D^c r) \gamma_{ij} D_c h ,
\nonumber \\
&&
    \bar{\nabla}_z h_{AB} = \partial_z h_{AB} .
\end{eqnarray}

\subsubsection{Perturbed Ricci tensor}

\begin{eqnarray}
& 2\delta {\bar R}_{ab}= 
& -\square h_{ab}+D_aD_ch^c_b+D_bD_ch^c_a
\nonumber
\\
&& +n\frac{D^cr}{r}(-D_ch_{ab}+D_ah_{cb}+D_bh_{ca})
\nonumber\\
&&
+{}^{(2)}\!R^c_ah_{cb}+{}^{(2)}\!R^c_bh_{ca}-2\,{}^{(2)}\!R_{acbd} 
h^{cd}-\frac{1}{r^2}\hat\triangle h_{ab}
\nonumber\\
&& +\frac{1}{r^2}(D_a\hat D^ih_{bi}+D_b\hat D^ih_{ai})
-\frac{D_br}{r^3}D_a h_{ij}\gamma^{ij}
-\frac{D_ar}{r^3}D_b h_{ij}\gamma^{ij}
\nonumber\\
&& +\frac{4}{r^4}D_arD_br h_{ij}\gamma^{ij}-D_aD_b h 
- \partial_z^2 h_{ab} ,
\end{eqnarray}

\begin{eqnarray}
& 2\delta {\bar R}_{ai}= 
& \hat D_iD_b h^b_a+\frac{n-2}{r}D^br \hat D_i h_{ab} \nonumber\\
&&-r\,\square\left(\frac{1}{r}h_{ai}\right)
-\frac{n}{r}D^br D_b h_{ai}
-D_ar D_b\left(\frac{1}{r}h^b_i\right)
\nonumber\\
&& +\frac{n+1}{r}D^br D_ah_{bi}
+ rD_a D_b\left(\frac{1}{r}h^b_i\right) 
\nonumber\\
&& +\left[(n+1)\frac{(Dr)^2}{r^2}+(n-1)\frac{K-(Dr)^2}{r^2}
-\frac{\square r}{r}\right]h_{ia}
\nonumber\\
&& + \frac{1}{r^2}D^br D_ar h_{bi}
+(n+1)rD_a\left(\frac{1}{r^2}D^br\right)h_{bi} \nonumber\\
&& -\frac{n+2}{r}D_aD^br h_{ib}
+ {}^{(2)}\!R^b_a h_{bi}-\frac{1}{r^2}\hat\triangle h_{ai}
+\frac{1}{r^2}\hat D_i\hat D^j h_{aj}
\nonumber\\
&& +rD_a\left(\frac{1}{r^3}\hat D^jh_{ji}\right)
+\frac{1}{r^3}D_ar \hat D^j h_{ji}
-\frac{1}{r^3}D_ar \hat D_i h_{jk}\gamma^{jk} 
\cr
&& -rD_a\left(\frac{1}{r}\hat D_i h\right)
- \partial_z^2 h_{ai} ,
\end{eqnarray}

\begin{eqnarray}
&2\delta {\bar R}_{ij}= & 
\left[2rD^ar D_bh^b_a+2(n-1)D^arD^br h_{ab}+2rD^aD^br 
h_{ab}\right]\gamma_{ij}
\nonumber\\
&& 
+ r \hat D_i D_a\left(\frac{1}{r}h^a_j\right)
+ r \hat D_j D_a\left(\frac{1}{r}h^a_i\right) 
\nonumber\\
&& + (n-1)\frac{D^ar}{r}(\hat D_ih_{aj}+\hat D_jh_{ai})
   + 2\frac{D^ar}{r}\hat D^k h_{ka} \gamma_{ij}
\nonumber\\
&&- r^2\,\square \left(\frac{1}{r^2}h_{ij}\right)
  - n \frac{D^ar}{r}D_ah_{ij}
  + \frac{1}{r^2}(\hat D_i\hat D^k h_{kj}+\hat D_j\hat D^k h_{ki})
\nonumber\\
&&-\frac{1}{r^2}\hat\triangle h_{ij}
+2\left[(n-1)\frac{K}{r^2}+2\frac{(Dr)^2}{r^2}-\frac{\square r}{r}
\right]h_{ij} 
\nonumber\\
&&
-2(\gamma^{kl}h_{kl}\gamma_{ij}-h_{ij})\frac{K-(Dr)^2}{r^2}-2\frac{(D
r)^2}{r^2}\gamma_{ij}\gamma^{kl}h_{kl} 
\nonumber\\
&& - \hat D_i\hat D_j h- rD^arD_a h \gamma_{ij}
   - \partial_z^2 h_{ij} ,
\label{eq:delta R_ij}
\end{eqnarray}

\begin{eqnarray}
 &\delta {\bar R} = & 
   D_aD_b h^{ab}+\frac{2n}{r}D^a rD^b h_{ab} 
\nonumber\\
&&
 + \left(-{}^{(2)}\!R^{ab}+\frac{2n}{r}D^aD^br 
            + \frac{n(n-1)}{r^2}D^ar D^b r\right)h_{ab}
\nonumber\\
&& + \frac{2}{r^2}D_a\hat D^i h_i^a
   + 2(n-1)\frac{D^ar}{r^3}\hat D^i h_{ai} 
\nonumber\\
&& + \frac{1}{r^4}\hat D^i \hat D^jh_{ij}
   - \frac{D^ar}{r^3}D_a h_{ij}\gamma^{ij}
   - \frac{1}{r^2}\left[(n-1)\frac{K}{r^2}-2\frac{(Dr)^2}{r^2}
   \right]h_{ij} \gamma^{ij}
\nonumber\\
&& - \square h - n\frac{D^ar}{r}D_a h
   - \frac{1}{r^2}\hat\triangle h
   - \partial_z^2 h ,
\end{eqnarray}

$\delta\bar{R}_{A z}$ components are 
\begin{eqnarray}
& 2\delta {\bar R}_{az}= 
& \partial_z \Bigl\{
    -D_a h + D_b h^b_a + \frac{1}{r^2} \hat{D}^j h_{ja}
    + n \frac{D^c r}{r} h_{ac} - \frac{D_a r}{r^3} (h_{ij}\gamma^{ij}) 
    \Bigr\}
\label{eq:delta R_az}
\\
& 2\delta {\bar R}_{iz}= 
& \partial_z \Bigl\{
    - \hat{D}_i h + D_b h^b_i + \frac{1}{r^2} \hat{D}^j h_{ji}
    + n \frac{D^c r}{r} h_{ci}  
    \Bigr\}
\label{eq:delta R_iz}
\\
& 2\delta {\bar R}_{zz}= 
& -  \partial_z^2  h
\label{eq:delta R_zz}
\end{eqnarray}

\section{Einstein Equations}
\label{app:Einstein Equations}

Einstein equations for scalar perturbations are summarized as follows.
From the components $\delta G_{i}^a$ and traceless part of $\delta G_{i}^j$ of the Einstein equations, we find the following equations:
\begin{eqnarray}
&& 
 k_S \left[\frac{1}{r^{n-2}}  
 D_b(r^{n-2}F_a^b) - r D_a \left(\frac{F^c_c}{r}\right)-2 (n-1)D_a F\right] 
 +  r f_{a,zz} = 0 ,  
\label{eq:delta G_ai} 
\\
&& 
  \frac{k_S^2}{2r^2} \Bigl[  2(n-2)F + F^c_c  \Bigr] + H_{T,zz} = 0 .
\label{eq:delta G_ij traceless} 
\end{eqnarray}
$\delta G_{ab}$ and $\delta G_{i}^i$ gives another two equations. 
\begin{eqnarray}
- \frac{1}{\mathbb{S}}2 \delta G_{ab}
&=& 
\square F_{ab}  - D_a D^c F_{bc} - D_b D^c F_{ac} + n \frac{D^c r}{r}(D_c F_{ab} - D_a F_{bc} - D_b F_{ac}) - {R^{(2)c}}_a F_{cb} - {R^{(2)c}}_b F_{ca}
\nonumber
\\
&&  + 2R^{(2)}_{acbd}F^{cd} - \frac{k_S^2}{r^2}F_{ab} + D_a D_b {F^c}_c + 2n\left(D_a D_b F + \frac{1}{r} D_a r D_b F + \frac{1}{r} D_b r D_a F\right)+f_{ab,zz}\cr
\nonumber
\\
&& 
 + g_{ab}
\Bigg[
  D^c D^d F_{cd} 
  + \frac{2n}{r}D^c r D^d F_{cd} 
  + \left( 
        \frac{2n}{r} D^c D^d r + n(n-1)\frac{D^c r D^d r}{r^2}-R^{(2)cd} 
    \right) F_{cd}
  - 2n {\square F}
\nonumber
\\
&&
  - \frac{2n(n+1)}{r}D^c r D_c F + 2(n-1) \frac{k_S^2-nK}{r^2}F 
  - \square F^c_c - \frac{n}{r} D^d r D_d  F^c_c + \frac{k_S^2}{r^2} F^c_c
  - f^c_{c,zz} - 2n H_{L,zz}\Bigg] 
\label{eq:scalar-dG_ab}
\end{eqnarray}

\begin{eqnarray}
- \frac{1}{\mathbb{S}} \frac{1}{n} \delta G^i_{i}
&=& \frac{1}{2} D^a D_b F_a^b
  - \frac{1}{2} \square F^a_a 
   + 
   \frac{(n-1)D^a r}{2r} 
   \left( 2D_b F_a^b  - D_a F^c_c \right)
   + \Bigg[
    (n-1)  \left(
        \frac{(n-2)}{2r^2} D^a r D^b r 
      + \frac{D^aD^b r}{r} 
\right)
      - \frac{R_{ab}^{(2)} }{2}
     \Bigg] F_{ab}
\cr 
&&
- (n-1) \square F
- \frac{n(n-1)}{r} D^a r D_a F 
+ 
  \frac{(n-1)}{2n r^2} 
    \Bigl[  2(n-2)  (k_S^2-nK)F 
           +    k_S^2 F^a_a
    \Bigr]
         - \frac{1}{2} f^a_{a,zz} - (n-1)H_{L,zz}  
\label{eq:delta G_ij trace}
\end{eqnarray}
Explicit equations from $\delta R_{zA}=0$ components are 
\begin{eqnarray}
2 \delta \bar{R}_{zz} &=& - \mathbb{S} ~ \partial_z^2 (f^c_c + 2n H_L ) = 0, 
\label{eq:R_zz-scalar component}
\\
2 \partial_z \delta \bar{R}_{az} 
&=& \mathbb{S} ~ \partial_z^2  
\left(
  D_cf^c_a + \frac{k_S}{r}f_a
  + n \frac{D^c r}{r} f_{ca}
  - 2n \frac{D_a r}{r} H_L
 \right) =0, 
\label{eq:R_za-scalar component}
\\
2 \partial_z \delta \bar{R}_{iz} 
&=& \mathbb{S}_{i} ~ \partial_z^2 
\left(
  D_c (r f^c)  + n(D^c r) f_c
- 2 k_S H_L 
 +  2 H_T  \left[ \frac{n-1}{n} \frac{k_S^2 - n K}{k_S}  \right]
\right) =0,
\label{eq:R_zi-scalar component}
\end{eqnarray}
where we have used (\ref{eq:R_zz-scalar component}) in (\ref{eq:R_za-scalar component}) and (\ref{eq:R_zi-scalar component}).

Let us try to rewrite Eqs. (\ref{eq:scalar-dG_ab}) and (\ref{eq:delta G_ij trace}).
Taking the trace of (\ref{eq:scalar-dG_ab}) and combining it with (\ref{eq:delta G_ij trace}), we can solve $f^c_{c,zz}$ and $H_{L,zz}$ in terms of $F$ and $F_{ab}$: 
\begin{eqnarray}
2(n+1) H_{L,zz}  
&=& 
 \square {F^c_c} 
    - D^a D_b F_a^b 
    - \frac{D^a r}{r}(D_a {F^b_b} - 2D_b F_a^b ) 
+ \biggl[
    2(n+1)\frac{D_a D_b r}{r} + (n-1)(n+2)\frac{D_a r D_b r}{r^2} 
    - R^{(2)}_{ab} 
  \biggr] F^{ab} 
\cr
&&  
   + \frac{k_S^2}{n r^2}{F^c_c}  
   - 2 \square F 
   - \frac{2n(n+1)}{r} D^a r D_a F 
   + \frac{2(n-1)(n+2)}{n}\frac{k_S^2-nK}{r^2} F 
\label{eq:sol H_L,zz}
\\
(n+1) {f^a}_{a,zz} 
 &=&     -  2n \square F^c_c
   +  2n D^a D^b F_{ab} 
   -  n(n-1) \frac{D^a r}{r} (D_a {F^b_b} - 2D_b F_a^b)
   + (n-1)\frac{k_S^2}{r^2} F^c_c
\cr
&&  -  2 \left[ R^{(2)}_{ab} + n(n-1) \frac{D_a r D_b r}{r^2}
                    \right] F^{ab}
    -  2n(n-1)  \square F 
    -  4(n-1) \frac{k_S^2-nK}{r^2} F 
\label{eq:sol f_a,zz}
\end{eqnarray}
Substituting these into (\ref{eq:scalar-dG_ab}) we obtain
\begin{eqnarray}
 && \square F_{ab} 
 - D_a D^c F_{bc} - D_b D^c F_{ac} 
 + n \frac{D^c r}{r}(D_c F_{ab} - D_a F_{bc} -  D_b F_{ac}) 
 - R^{(2)}_{ca} F^c_b - R^{(2)}_{cb} F^c_a
\cr
&&\qquad 
 + 2R^{(2)}_{acbd}F^{cd} 
 - \frac{k_S^2}{r^2}F_{ab} + D_a D_b  F^c_c 
 + 2n\left(D_a D_b F + \frac{1}{r} D_a r D_b F + \frac{1}{r} D_b r D_a F\right) 
\cr
&&\qquad 
 + \frac{g_{ab}}{n+1}\Bigg[D^c D^d F_{cd} 
 - \frac{n}{r}D^c r(D_c {F^d_d} -2 D_d {F_c^d}) 
 + \left(R^{(2)}_{cd} + n(n-1) \frac{D_c r D_d r}{r^2}\right)F^{cd} 
\cr
&&\qquad 
 - 2n{\square F}-\frac{2n(n+1)}{r}D^c r D_c F 
 + 2(n-1) \frac{k_S^2-nK}{r^2}F 
 -  \square {F^c_c}  + \frac{k_S^2}{r^2}{F^c_c} \Bigg]
 + f_{ab,zz}
  =0
\label{eq:equation for ell=1}
\end{eqnarray}

So far, we have only used (\ref{eq:scalar-dG_ab}) and (\ref{eq:delta G_ij trace}).
From (\ref{eq:delta G_ai}) and (\ref{eq:delta G_ij traceless}), one can construct $X_a$ as
\begin{eqnarray}
\partial_z^2 X_a
    =
       - \frac{1}{r^{n-2}}D_b ( r^{n-2} F^b_a )
       + \frac{1}{2} D_a F_c^c + n D_a F
       + 2(n-2) \frac{D_a r}{r} F . 
\label{eq:def D_zz Za}
\end{eqnarray}
Then using this, we can rewritten (\ref{eq:equation for ell=1}) and (\ref{eq:sol H_L,zz}) in terms of $F_{ab}$ and $F$, resulting in
Eqs. (\ref{eq:D^aD^b F_ab}) and  (\ref{eq:Box F_ab}).

\bibliographystyle{apsrev} 



\end{document}